\documentclass[fleqn,usenatbib,useAMS]{mnras}

\usepackage{newtxtext,newtxmath}

\usepackage[T1]{fontenc}

\DeclareRobustCommand{\VAN}[3]{#2}
\let\VANthebibliography\thebibliography
\def\thebibliography{\DeclareRobustCommand{\VAN}[3]{##3}\VANthebibliography}

\usepackage{graphicx}	
\usepackage{amsmath}	
\usepackage{orcidlink}

\newcommand{\msun}{{\rm M}_{\sun}}

\title[Magnetic ultramassive white dwarfs]{Magnetic field breakout in ultramassive crystallizing white dwarfs}

\author[D. Blatman and S. Ginzburg]{
Daniel Blatman$^{\orcidlink{0009-0001-1957-8801}}$\thanks{E-mail: \href{mailto:daniel.blatman1@mail.huji.ac.il}{daniel.blatman1@mail.huji.ac.il}} and
Sivan Ginzburg$^{\orcidlink{0000-0002-3751-4553}}$\thanks{E-mail: \href{mailto:sivan.ginzburg@mail.huji.ac.il}{sivan.ginzburg@mail.huji.ac.il}}
\\
Racah Institute of Physics, The Hebrew University, Jerusalem 9190401, Israel
}

\date{Accepted XXX. Received YYY; in original form ZZZ}

\pubyear{\the\year{}}

\begin{document}
\label{firstpage}
\pagerange{\pageref{firstpage}--\pageref{lastpage}}
\maketitle

\begin{abstract}
Ultramassive white dwarfs with masses $M\gtrsim 1.1\,{\rm M}_{\sun}$ probe extreme physics near the Chandrasekhar limit. Despite the rapid increase in observations, it is still unclear how many harbour carbon--oxygen (CO) versus oxygen--neon (ONe) cores. The origin of these white dwarfs and their strong magnetic fields -- single stellar evolution or a stellar merger -- is another open question. The steep mass--radius relation of the relativistic ultramassive white dwarfs shortens their crystallization time $t_{\rm cryst}$, such that the recently proposed crystallization dynamo mechanism may present an alternative to mergers in explaining the early appearance of magnetism in the observed population. However, the magnetic diffusion time from the convective dynamo to the white dwarf's surface delays the magnetic field's breakout time $t_{\rm break}>t_{\rm cryst}$. We compute $t_{\rm break}(M)$ for CO and ONe ultramassive white dwarfs and compare it to the local 40 pc volume-limited sample. We find that the breakout time from CO cores is too long to account for the observations. ONe crystallization dynamos remain a viable option, but their surrounding non-convective envelopes comprise only a few per cent of the total mass, such that $t_{\rm break}$ is highly sensitive to the details of stellar evolution.
\end{abstract}




\section{Introduction}

When a main sequence star exhausts the hydrogen fuel at its centre, the helium ash of nuclear burning builds up a degenerate core while the hydrogen envelope expands on the red giant branch \citep[see][for a review of standard stellar evolution]{Kippenhahn2012}. Hydrogen burning continues in a thin shell around the degenerate helium core, gradually increasing the core's mass. When the core exceeds a mass of $\approx 0.5\,\msun$, the helium ignites, eventually producing a carbon--oxygen (CO) degenerate core. Helium-shell burning continues to increase the underlying core's mass as the hydrogen envelope expands on the analogous asymptotic giant branch (AGB). Strong stellar winds during the AGB phase remove the hydrogen envelope, truncating the core's growth at some point, leaving behind a degenerate CO white dwarf. However, an initially massive enough star may grow a CO core to a mass of $\approx 1.1\,\msun$, when carbon ignites. Such stars enter another analogous super-AGB phase, with a degenerate oxygen--neon (ONe) core surrounded by a carbon-burning shell. If the ONe core approaches the Chandrasekhar mass ($\approx 1.4\,\msun$), it collapses in a supernova explosion. However, for a narrow range of initial stellar masses, winds during the super-AGB phase truncate the ONe core's growth -- potentially leaving behind an ONe white dwarf \citep{Rakavy1967,Murai1968, GarciaBerro1997,Siess2006,Siess2010,Kippenhahn2012,Doherty2015,Doherty2017}.

Although ultramassive ($\gtrsim 1.1\,\msun$) white dwarfs are now being routinely discovered \citep{Thejill1990,Bergeron1992,Schmidt1992,VennesKawka2008,Hermes2013,Cummings2016,Kilic2021}, both the formation and composition of these stars are still uncertain. Whereas some ultramassive white dwarfs seem to be consistent with the single stellar evolution described above \citep{Miller2022,Miller2023,Kilic2023single}, others may have formed through a merger of two lower mass white dwarfs \citep{Hollands2020,Pshirkov2020,Caiazzo2021,Kilic2024}; the contribution of these two channels to the overall population appears to be similar  \citep{Temmink2020,Fleury2022,Kilic2023}. The critical mass $\approx 1.1\,\msun$ distinguishing ONe from CO white dwarfs is sensitive to rotation \citep{Dominguez1996} and stellar winds \citep{Althaus2021}. Whether or not this carbon-ignition mass is different in white dwarf mergers is still under debate \citep{Schwab2021,Wu2022,Shen2023}. Observationally, a recently detected clustering in the \textit{Gaia} data (the Q branch) appears to favour CO, rather than ONe, cores for at least some ultramassive white dwarfs \citep{Cheng2019,Tremblay2019,Bauer2020,Blouin2021,Camisassa2021,Bedard2024}. On the other hand, the high measured neon abundance in several novae indicates that some ONe white dwarfs also exist \citep{TruranLivio1986,LivioTruran1994}.

Many ultramassive white dwarfs are strongly magnetized, with measured magnetic fields typically exceeding 1 MG \citep{BagnuloLandstreet2022}. These strong fields are often interpreted as an indication of a double white dwarf merger, with fast differential rotation driving a magnetic dynamo \citep{GarciaBerro2012}. However, fields of similar strength are also observed for many lower mass CO white dwarfs. One of the promising mechanisms to account for these magnetic CO white dwarfs is compositional convection during their crystallization \citep{Stevenson1980,Mochkovitch1983,Isern1997}. When white dwarfs cool down sufficiently, their interiors begin to crystallize from the inside out. The crystal phase is enriched in oxygen and depleted in carbon compared to the liquid phase, generating an unstable compositional gradient in the liquid layers above the crystal core. The ensuing convection may drive a magnetic dynamo on its own \citep{Isern2017} or transport primordial fossil fields closer to the white dwarf's surface. While the magnitude of such crystallization dynamos is still uncertain \citep{Schreiber2021Nat,Ginzburg2022,Fuentes2023,Fuentes2024,Castro-Tapia2024,MontgomeryDunlap2024}, their timing can be compared to the observed magnetic white dwarf ages. Specifically, \citet{BagnuloLandstreet2022} found that in a volume-limited sample strong magnetic fields appear predominantly after the onset of crystallization,  emphasizing this channel's potential contribution. In \citet{BlatmanGinzburg2024} we conducted a more careful analysis, by considering the magnetic field's breakout time to the surface rather than the crystallization time. At the early stages of crystallization, the magnetic field is trapped inside the convective region above the crystal core. Only once a significant (and mass dependent) portion of the white dwarf has crystallized, convection approaches the white dwarf's envelope sufficiently and the field reaches the surface by magnetic diffusion. This breakout process delays the emergence of magnetism by up to a few Gyr, depending on the white dwarf's mass.

The two-component phase diagrams of CO and ONe are qualitatively similar, with the heavier component (i.e. oxygen for CO mixtures and neon for ONe mixtures) being more abundant in the crystal phase than in the liquid phase \citep{MedinCumming2010,BlouinDaligault2021CO,BlouinDaligault2021ONe}. This similarity implies that ONe white dwarfs may also harbour crystallization dynamos. \citet{Camisassa2022} proposed to exploit the difference in crystallization times between CO and ONe white dwarfs with the same mass \citep[see also][]{Bauer2020,Camisassa2021,Camisassa2022first} to identify the composition of some magnetic ultramassive white dwarfs, assuming that their fields are generated by a crystallization dynamo.

Here, we compute the breakout time of magnetic fields to the surface of both CO and ONe ultramassive white dwarfs, extending our previous work on lower mass CO white dwarfs \citep{BlatmanGinzburg2024}. This breakout time may be significantly longer than the crystallization time -- delaying the observed appearance of magnetism. The magnetic breakout times may be used to identify the formation channel (single star or merger), composition (CO or ONe, and the carbon, oxygen, and neon profiles), and magnetization mechanism (crystallization or merger) of the observed ultramassive white dwarf population.    

The remainder of this letter is organized as follows. In Section \ref{sec:anal_cryst} we derive an analytical model for the crystallization time of ultramassive white dwarfs. In Section \ref{sec:numerical} we compute the crystallization and breakout times numerically. We discuss the implications on the observed population in Section \ref{sec:obs}, and summarize our conclusions in Section \ref{sec:summary}.

\section{Ultramassive white dwarf crystallization}\label{sec:anal_cryst}

In \citet{BlatmanGinzburg2024} we used the \citet{Mestel1952} white dwarf cooling theory to derive an analytical relation $t_{\rm cryst}\propto M^{-5/3}$ between the white dwarf's mass $M$ and its crystallization time $t_{\rm cryst}$. Here, we extend this analytical theory to ultramassive white dwarfs that approach the \citet{Chandrasekhar1931,Chandrasekhar1931MNRAS,Chandrasekhar1935} mass $M_{\rm ch}$.

The pressure $P$ in the interior of ultramassive white dwarfs is given by relativistic degenerate electrons, approaching $P\sim hcn_e^{4/3}$, where $h$ is Planck's constant, $c$ is the speed of light and $n_e$ is the electron number density. Further from the centre, $n_e$ decreases and the degenerate electrons become non-relativistic with $P\sim h^2m_e^{-1}n_e^{5/3}$, where $m_e$ is the electron's mass \citep[see][for details and omitted numerical coefficients]{Hansen2004}. Thanks to the heat conductivity of degenerate electrons, the entire degenerate region is approximately isothermal, with a temperature $T_{\rm c}$. As $n_e$ decreases even further close to the white dwarf's edge, the degeneracy is lifted when the ideal gas pressure becomes comparable to the degeneracy pressure
\begin{equation}\label{eq:deg_lift}
P\sim n_e k_{\rm B} T_{\rm c}\sim \frac{h^2}{m_e}n_e^{5/3},    
\end{equation}
where $k_{\rm B}$ is Boltzmann's constant. Equation \eqref{eq:deg_lift} indicates that the degeneracy is lifted at $n_e<(m_ec/h)^3$ as long as $k_{\rm B}T_{\rm c}<m_ec^2$ \citep[white dwarfs crystallize at much lower central temperatures; see][]{VanHorn1968}, such that the degeneracy is lifted in the non-relativistic outer layers even for ultramassive white dwarfs.

The white dwarf's cooling is regulated by photon diffusion through the geometrically thin non-degenerate envelope. Using equation \eqref{eq:deg_lift}, the density $\rho$ at the base of this envelope scales as $\rho\propto n_e\propto T_c^{3/2}$ and the pressure scales as $P\propto T_c^{5/2}$. The envelope's optical depth is given by $\tau\sim P\kappa/g$, where the opacity $\kappa\propto\rho T_c^{-7/2}$ follows Kramers' law and the surface gravity $g=GMR^{-2}$, with $G$ denoting the gravitational constant and $R$ the white dwarf's radius. We find the luminosity $L$ by applying the diffusion equation
\begin{equation}
    L\sim\frac{4\upi R^2\sigma T_c^4}{\tau}\propto\frac{R^2 T_c^4}{R^2 M^{-1} T_c^{1/2}}=MT_c^{7/2},
\end{equation}
where $\sigma$ is the Stefan--Boltzmann constant, and the radius $R$ cancels out \citep{Mestel1952}. The white dwarf's heat capacity is dominated by the non-degenerate ions, such that the cooling time down to a given central temperature scales as
\begin{equation}\label{eq:cooling_time}
t\propto \frac{M k_{\rm B}T_c}{L}\propto T_c^{-5/2},
\end{equation}
where the mass $M$ cancels out as well.

As the white dwarf cools and $T_{\rm c}$ decreases, its interior begins to crystallize when the plasma coupling parameter at the centre
\begin{equation}\label{eq:Gamma}
\Gamma\propto\frac{\rho_{\rm c}^{1/3}}{T_{\rm c}}\sim \frac{M^{1/3}}{RT_{\rm c}}   
\end{equation}
exceeds a critical value $\Gamma\approx 200$ \citep{VanHorn1968,PotekhinChabrier2000,Bauer2020,Jermyn2021}, where the central density is given by $\rho_c\sim M R^{-3}$. Using equations \eqref{eq:cooling_time} and \eqref{eq:Gamma}, the crystallization time scales as
\begin{equation}\label{eq:t_cryst_mr}
t_{\rm cryst}\propto T_{\rm c}^{-5/2}\propto M^{-5/6}R^{5/2}.    
\end{equation}
We adopt the approximate white dwarf mass--radius relation from \citet{Hansen2004}\footnote{Private communication by Eggleton (1982) in \citet{TruranLivio1986}.}
\begin{equation}\label{eq:mass_rad}
   R\propto M^{-1/3}\left[1-\left(\frac{M}{M_{\rm ch}}\right)^{4/3}\right]^{1/2}, 
\end{equation}
which is a fit to numerical integration of the hydrostatic equilibrium equation with the accurate $P(n_e)$, properly transitioning from $P\propto n_e^{5/3}$ to $P\propto n_e^{4/3}$. In particular, equation \eqref{eq:mass_rad} reproduces the non-relativistic limit $R\propto M^{-1/3}$ for $M\ll M_{\rm ch}$ and the ultra-relativistic limit $R\to 0$ for $M\to M_{\rm ch}$.\footnote{General relativity becomes important for white dwarfs with $M\gtrsim 1.3\,\msun$ \citep{Carvalho2018,Althaus2022,Althaus2023}, which are beyond our scope.} By plugging this relation in equation \eqref{eq:t_cryst_mr}, we find
\begin{equation}\label{eq:t_cryst}
t_{\rm cryst}(M)\propto M^{-5/3}\left[1-\left(\frac{M}{M_{\rm ch}}\right)^{4/3}\right]^{5/4},     
\end{equation}
which reduces to $t_{\rm cryst}\propto M^{-5/3}$ for $M\ll M_{\rm ch}$ \citep{BlatmanGinzburg2024}, and approaches $t_{\rm cryst}\to 0$ for ultramassive white dwarfs with $M\to M_{\rm ch}$. In the following Section \ref{sec:numerical}, we demonstrate that our analytical equation \eqref{eq:t_cryst} agrees well with a more accurate numerical calculation of $t_{\rm cryst}(M)$. 

\section{Numerical results}\label{sec:numerical}

We use the stellar evolution code \textsc{mesa}, version r24.03.1 \citep{Paxton2011,Paxton2013,Paxton2015,Paxton2018,Paxton2019,Jermyn2023} to construct initial CO and ONe white dwarf models using two different methods. In the first method, CO white dwarfs with $M\leq 1.05\,\msun$ and ONe white dwarfs with $1.1\,\msun\leq M\leq 1.2\,\msun$ are evolved from pre-main sequence progenitors with different initial masses using the test suites \texttt{make\_co\_wd} and \texttt{make\_o\_ne\_wd}.\footnote{We use the nominal \citet{DeBoer2017} $^{12}{\rm C}(\alpha,\gamma)^{16}{\rm O}$ nuclear reaction rates \citep[expanded by][]{Mehta2022}, whereas in \citet{BlatmanGinzburg2024} both the \citet{Kunz2002} and the \citet{Mehta2022} rates were used.} In the second method, we use the \texttt{relax\_mass\_scale} procedure to re-scale the $1.05\,\msun$ CO white dwarf and the $1.15\,\msun$ and $1.18\,\msun$ ONe white dwarfs (created using the first method) to different total masses $M$, while preserving the chemical profile as a function of the relative mass coordinate $m/M$. We then compute the cooling and crystallization of all the initial white dwarf models (created by both methods) using the test suite \texttt{wd\_cool\_0.6M}.

\begin{figure}
\includegraphics[width=\columnwidth]{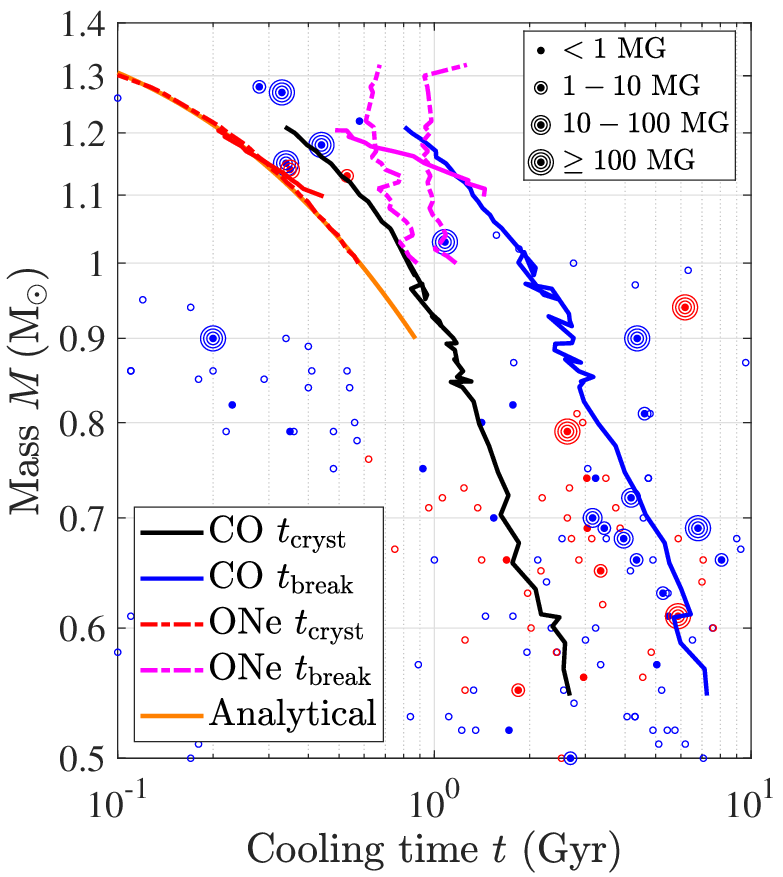}
\caption{Crystallization and magnetic field breakout times for CO and ONe white dwarfs. CO white dwarfs with mass $M\leq 1.05\,\msun$ are evolved from main sequence progenitors, whereas more massive CO white dwarfs are scaled versions of the $M=1.05\,\msun$ model. Similarly, we construct ONe white dwarfs with $1.1\,\msun\leq M\leq 1.2\,\msun$ through stellar evolution (solid red and magenta lines), and scale the $1.15\,\msun$ and $1.18\,\msun$ models to both lower and higher masses (dashed lines). Whereas the ONe $t_{\rm cryst}(M)$ is robust and fits our analytical equation \eqref{eq:t_cryst}, the ONe $t_{\rm break}(M)$ is sensitive to the assumed O/Ne profile, and therefore to the mass of the source model (see Fig. \ref{fig:profiles}). The observed white dwarfs (blue for hydrogen atmospheres and red for helium) are from the volume-limited sample of \citet{BagnuloLandstreet2022}. Empty dots represent non-magnetic white dwarfs and filled dots mark magnetic ones, with the strength of the measured magnetic field indicated by the number of surrounding circles.}
\label{fig:obs_ultra}
\end{figure}

The onset of crystallization is determined by the Skye equation of state \citep{Jermyn2021} and plotted in Fig. \ref{fig:obs_ultra}. As found in previous studies \citep{Bauer2020,Camisassa2021,Camisassa2022first,Camisassa2022}, ONe white dwarfs crystallize before CO white dwarfs with the same mass (due to the different ion electrical charge). For the most part, our ONe crystallization curve $t_{\rm cryst}(M)$ is independent of the method used to generate the initial white dwarf model (or the source model used for mass re-scaling), and it fits well our analytical equation \eqref{eq:t_cryst} with $M_{\rm ch}=1.46\,\msun$ \citep{Hansen2004,Kippenhahn2012}. Specifically, as indicated by equation \eqref{eq:t_cryst}, the $t_{\rm cryst}\propto M^{-5/3}$ power law bends for ultramassive white dwarfs, such that they crystallize much earlier than lower-mass stars. 

During crystallization, the phase separation of CO or ONe into a crystal (enriched with the heavier element) and a liquid (enriched with the lighter element) is computed using the \texttt{phase\_separation} scheme \citep{Bauer2023}, according to the \citet{BlouinDaligault2021CO,BlouinDaligault2021ONe} phase diagrams. This scheme determines the outer radius of the convective region iteratively until a stable composition profile is obtained. We calculate the magnetic diffusion time from this radius to the white dwarf's surface and determine the breakout time $t_{\rm break}$ similarly to \citet{BlatmanGinzburg2024}. 

\begin{figure}
\includegraphics[width=\columnwidth]{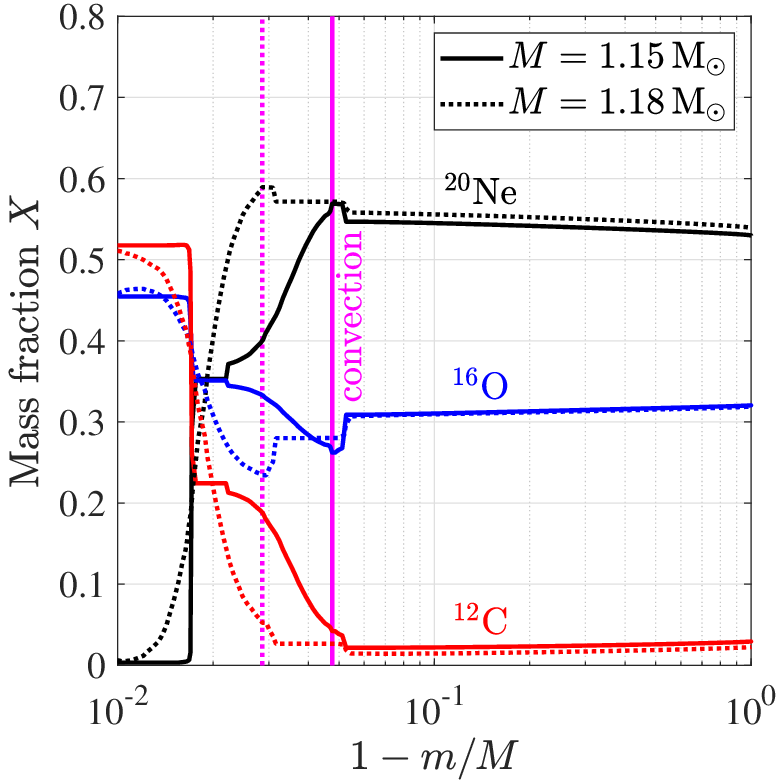}
\caption{Initial neon, oxygen, and carbon abundances $X$ as a function of the relative mass coordinate $m/M$ for the two ONe source models used in Fig. \ref{fig:obs_ultra}, where these profiles are scaled to different white dwarf masses $1.0\,\msun<M<1.3\,\msun$ while preserving $X(m/M)$. The crystallization of the white dwarf's core drives convection, which extends up to the compositional transition region (vertical magenta lines). From there, magnetic fields diffuse to the surface. This diffusive envelope is almost twice as massive in models derived from $M=1.15\,\msun$ compared to $M=1.18\,\msun$, roughly doubling the diffusion time $t_{\rm diff}=t_{\rm break}-t_{\rm cryst}$, as seen in Fig. \ref{fig:obs_ultra}.}
\label{fig:profiles}
\end{figure}

Fig. \ref{fig:obs_ultra} demonstrates that unlike $t_{\rm cryst}$, the ONe breakout time $t_{\rm break}$ is highly sensitive to the assumed oxygen and neon profiles. Even a mild difference in the stellar evolution (which we parametrize by applying \texttt{relax\_mass\_scale} to source profiles with slightly different masses: $1.15\,\msun$ and $1.18\,\msun$) leads to a dramatic effect on $t_{\rm break}$. The origin of this sensitivity is elucidated in Fig. \ref{fig:profiles}, where we plot the chemical profiles. Similarly to \citet{BlatmanGinzburg2024}, the crystallization-driven convection zone encompasses the approximately uniform inner core of the white dwarf, stopping at the steep compositional gradient above it (vertical magenta lines in Fig. \ref{fig:profiles}). Unlike in the lower-mass CO white dwarfs studied in \citet{BlatmanGinzburg2024}, this convective boundary advances only slightly between $t_{\rm cryst}$ and $t_{\rm break}$. In our ONe white dwarfs, the outer convectively stable envelope contains only a few per cent of the total mass \citep[see also][]{Camisassa2022}, with the exact value depending critically on the details of stellar evolution. The time it takes magnetic fields to diffuse through this thin envelope to reach the surface is comparable to $t_{\rm cryst}$ or even longer, leading to large variations in the breakout time. 

In fact, the uncertainty in $t_{\rm break}$ could be even larger than illustrated in Fig. \ref{fig:obs_ultra}. Chemical profiles of ONe white dwarfs resulting from state-of-the-art stellar evolution calculations by various groups differ substantially from one another \citep{Lauffer2018,Camisassa2019,Bauer2020}. These evolutionary calculations are sensitive to uncertain parameters such as convective overshooting and super-AGB winds. The chemical profiles of white dwarf merger remnants are even less constrained \citep{Schwab2021,Schwab2021_most}. 

\section{Observations}\label{sec:obs}

We populate Fig. \ref{fig:obs_ultra} with the volume-limited sample of \citet{BagnuloLandstreet2022}. They interpreted the early appearance of strong magnetic fields in ultramassive white dwarfs as an indication of a merger origin. While we also find that some ultramassive magnetic white dwarfs are indeed younger than the CO $t_{\rm cryst}(M)$ line, all of these stars are older than our ONe $t_{\rm cryst}(M)$ curve. Judging by the crystallization time alone, we may therefore conclude that these ultramassive white dwarfs could have been magnetized by a crystallization dynamo, regardless of whether they are the products of single stellar evolution or a merger.\footnote{The cooling time (which is reset by mergers), and not the actual age, is the relevant quantity throughout this letter.} Similarly to \citet{Camisassa2022}, we find that a significant portion of the ultramassive magnetic white dwarfs lie in between the ONe and CO crystallization curves (2 out of 6 ultramassive white dwarfs with fields above a MG). If these stars were magnetized by a crystallization dynamo, then we may determine that they harbour ONe cores. Alternatively, if these stars harbour CO cores, then we may rule out a crystallization dynamo as the source of their magnetism.

Considering the magnetic diffusion time from the convective dynamo to the surface complicates this analysis. Our CO $t_{\rm break}(M)$ times appear to be too long to account for the observed magnetic ultramassive white dwarfs. While our ONe breakout times are generally shorter -- the $1.2\,\msun$ full stellar evolution model approaches the observations -- the ONe $t_{\rm break}(M)$ curve critically depends on the assumed O/Ne profile within the star (see Section \ref{sec:numerical}). It therefore remains unclear whether ONe crystallization could explain ultramassive white dwarf magnetism. 

\begin{figure}
\includegraphics[width=\columnwidth]{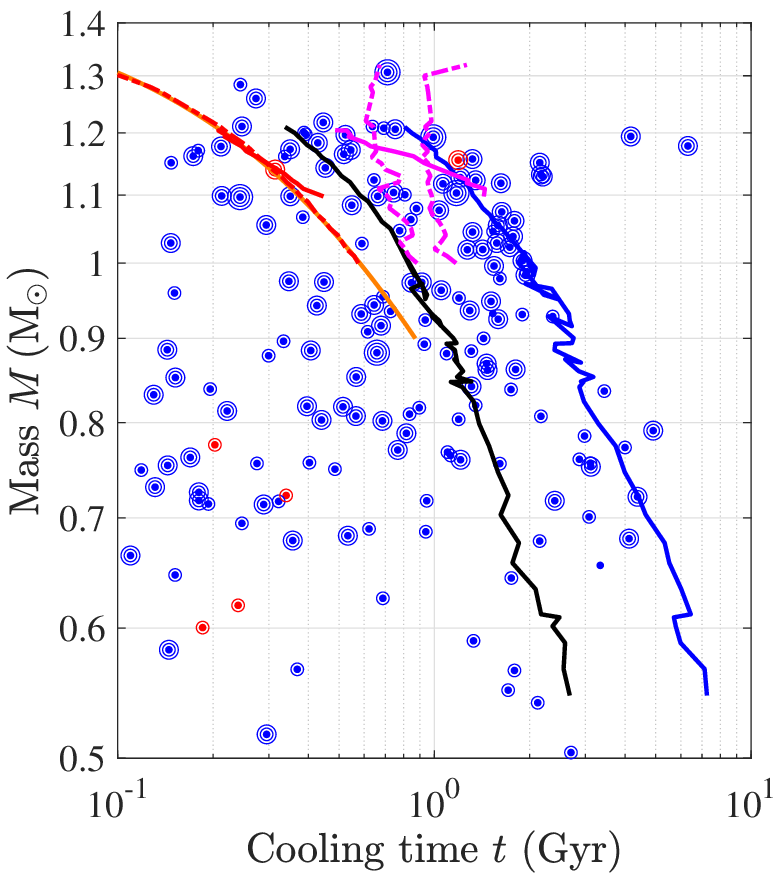}
\caption{Same as Fig. \ref{fig:obs_ultra}, but showing the \citet{Hardy_a,Hardy_b} sample of magnetic white dwarfs from the Montreal White Dwarf Database \citep{Dufour2017}. 
As in Fig. \ref{fig:obs_ultra}, blue  markers indicate hydrogen atmospheres \citep{Hardy_a}, and red markers indicate helium atmospheres \citep{Hardy_b}.
Unlike Fig. \ref{fig:obs_ultra}, this sample is not volume-limited, such that young white dwarfs are over-represented.}
\label{fig:hardy}
\end{figure}

The strength of volume-limited samples such as \citet{BagnuloLandstreet2022} is the absence of observational bias, but this comes at the cost of a small sample size. For a complementary picture, we also compare our crystallization and breakout curves to the \citet{Hardy_a,Hardy_b} sample in Fig. \ref{fig:hardy}. This sample contains almost all the white dwarfs from the Montreal White Dwarf Database\footnote{\url{https://www.montrealwhitedwarfdatabase.org}} \citep{Dufour2017} that show clear signs of line splitting (magnetic fields $B\gtrsim 1\textrm{ MG}$). This sample is not volume-complete, and as such suffers from various observational biases. In particular, most of the white dwarfs in the sample are drawn from magnitude-limited surveys, in which young stars (i.e. short cooling times) are over-represented. Nevertheless, Fig. \ref{fig:hardy} demonstrates that at least some ultramassive white dwarfs (as well as lower mass ones) acquire strong magnetic fields before our ONe and CO crystallization times, implying that other magnetization channels must exist.

\section{Conclusions}\label{sec:summary}

Ultramassive white dwarfs ($M\gtrsim 1.1\,\msun$) have been studied extensively over the last several years both theoretically and observationally. It is still unclear, however, how many of these white dwarfs formed through single stellar evolution and how many through white dwarf mergers, which of them harbour CO vs. ONe cores, and what is the origin of their strong magnetic fields. \citet{BagnuloLandstreet2022} argued that the early appearance of strong magnetic fields in ultramassive white dwarfs much younger than a Gyr indicates formation and magnetization by a merger. \citet{Camisassa2022}, on the other hand, pointed out that at least some of these ultramassive white dwarfs could have been magnetized by a convective crystallization dynamo instead \citep{Isern2017}. They suggested that the difference between CO and ONe crystallization times could be used to determine the core composition of these stars. In this letter, we reanalysed the timing of CO and ONe crystallization dynamos by considering the magnetic diffusion time from the outer edge of the convective dynamo to the white dwarf's surface, using the newly implemented phase separation scheme in \textsc{mesa} \citep{Bauer2023}. 

The steepening mass--radius relation of ultramassive white dwarfs (due to the relativistic degenerate electrons) dramatically shortens their crystallization time $t_{\rm cryst}$ as a function of mass $M$. The crystallization time is shortened further in ONe white dwarfs -- compared to CO white dwarfs with the same mass -- due to the stronger Coulomb interactions of the ions \citep{Bauer2020,Camisassa2021,Camisassa2022first,Camisassa2022}. As a result, the entire ultramassive white dwarf population within the local 40 pc volume-limited sample \citep{BagnuloLandstreet2022} is consistent with an ONe crystallization dynamo -- if the diffusion time to the surface is ignored, i.e. if other mechanisms can transport the magnetic field to the surface faster \citep{CharbonneauMacGregor2001,MacGregorCassinelli2003,MacDonaldMullan2004}. About a third of the ultramassive magnetic white dwarfs (notwithstanding the small-number statistics) lie in between the ONe and the CO crystallization curves. For these stars, a CO crystallization dynamo can be safely ruled out, i.e. they either harbour ONe cores or were magnetized through a merger \citep[see also][]{Camisassa2022}.

However, accounting for the magnetic diffusion time to the surface delays the emergence of magnetic fields from crystallization dynamos. We find that this delay rules out a CO crystallization dynamo scenario for the entire ultramassive white dwarf population in the sample, at least up to $M=1.2\,\msun$. As for ONe ultramassive white dwarfs, the picture is more complicated. The diffusive envelope in these stars is only a few per cent of the total mass, and it is very sensitive to small variations in the preceding stellar evolution. Therefore, while we find that ONe magnetic breakout times $t_{\rm break}(M)$ are generally shorter than for CO white dwarfs with the same mass, it remains unclear whether they are short enough to explain the observations.

We conclude that magnetic breakout from ONe -- but not CO -- crystallization dynamos presents a viable alternative to mergers in explaining some observed young magnetic ultramassive white dwarfs. If this is the dominant magnetization channel, then by extrapolating our results from Figs \ref{fig:obs_ultra} and \ref{fig:profiles}, the non-convective envelope of these stars must be smaller than about 1--2 per cent of their total mass. This stringent constraint on the chemical profile may in turn constrain the stellar evolution; in particular, whether these ultramassive white dwarfs formed from a single star or through a (non-magnetizing) merger.
We emphasize that an ONe crystallization dynamo cannot be the sole magnetization mechanism, because some magnetic white dwarfs in magnitude-limited surveys are younger than our ONe $t_{\rm cryst}(M)$; see Fig. \ref{fig:hardy}.

During the revision of this letter, \citet{Castro-Tapia2024_new} published an analysis of magnetic diffusion in CO white dwarfs with $M\leq 1.0\,\msun$, focusing on the evolution of the surface magnetic field strength $B(t)$. It could be interesting to extend this analysis to the ultramassive white dwarfs studied here. Specifically, we find that the initial convection zone in ONe white dwarfs extends to $\approx 0.8 R$, maximizing the surface $B$ according to \citet{Castro-Tapia2024_new}.

\section*{Acknowledgements}

We thank Ana Antonini, Evan Bauer, Simon Blouin, Jim Fuller, and Na'ama Hallakoun for motivating discussions. We also thank the organizers of the `Current challenges in the physics of white dwarf stars' workshop in Santa Fe (March 2024) for an excellent meeting where many of the ideas presented here were discussed. Finally, we thank the reviewer for a helpful report which improved the letter. We acknowledge support from the Israel Ministry of Innovation, Science, and Technology (grant No. 1001572596), and from the United States -- Israel Binational Science Foundation (BSF; grant No. 2022175).
\texttt{MesaScript} \citep{bill_wolf_mesascript} was used to automate some of the steps in this work.

\section*{Data Availability}

The \textsc{mesa} input files required to reproduce our results are available at \url{https://zenodo.org/doi/10.5281/zenodo.11583092}.



\bibliographystyle{mnras}
\input{massive.bbl}






\bsp	
\label{lastpage}
\end{document}